\documentclass[10pt]{article}

\usepackage[T1]{fontenc}										
\usepackage{lmodern}												
\usepackage{microtype}											
\usepackage[LGRgreek,frenchmath]{mathastext}
\usepackage[mathscr]{eucal}                 
\usepackage[sans]{dsfont}										
\usepackage{bbold}													
\usepackage{bbm}														
\usepackage{graphicx}                      	
\usepackage{mdwlist}												
\usepackage{natbib}    											
\setcitestyle{authoryear}
\usepackage[dvipsnames]{xcolor}							
\usepackage[plainpages=false, pdfpagelabels, backref=page]{hyperref} 
	\hypersetup{
		colorlinks   = true,
		citecolor    = RoyalBlue, 
		linkcolor    = RubineRed, 
		urlcolor     = MidnightBlue 
	}
\usepackage[paperwidth=8.5in,paperheight=11.00in,top=1.00in, bottom=1.00in, left=0.75in, right=0.75in]{geometry}
\usepackage{mathtools}                      
\mathtoolsset{showonlyrefs=true}            
\usepackage{amsmath}
\usepackage{amssymb}
\usepackage{amsfonts}
\usepackage{amsthm}                         
\allowdisplaybreaks                         
\newtheoremstyle{plain}
  {}   				
  {}   				
  {\itshape}  
  {}       		
  {\mdseries\scshape} 
  {.}         
  { } 				
  {\thmname{#1}\thmnumber{ #2}\ifx#3\empty\else\ (#3)\fi}
\theoremstyle{plain}
\newtheorem{theorem}{\underline{Theorem}}
       	
\newtheorem{proposition}[theorem]{\underline{Proposition}}

\newtheoremstyle{definition}
  {}   				
  {}   				
  {}  				
  {}      		
  {\mdseries\scshape} 
  {.}         
  { } 				
  {\thmname{#1}\thmnumber{ #2}\ifx#3\empty\else\ (#3)\fi}
\theoremstyle{definition}
\newtheorem{definition}[theorem]{\underline{Definition}}

\newtheorem{remark}[theorem]{\underline{Remark}}
\newtheorem{assumption}[theorem]{\underline{Assumption}}
\usepackage{sectsty}
\allsectionsfont{\mdseries\scshape}

\usepackage{xcolor}
\usepackage{listings}
\definecolor{mma_purple}{rgb}{0.5, 0, 0.5}
\definecolor{mma_blue}{rgb}{0, 0, 1}
\definecolor{mma_green}{rgb}{0, 0.5, 0}
\definecolor{mma_gray}{rgb}{0.5, 0.5, 0.5}
\definecolor{mma_background}{rgb}{0.98, 0.98, 0.98}
\usepackage{pdfrender}
\newcommand{\heavytext}[1]{\textpdfrender{
  TextRenderingMode=FillStroke,
  LineWidth=.3pt,
}{#1}}

\renewcommand{\[}{\left[}

\newcommand\Cb{\mathds{C}}
\newcommand\Eb{\mathds{E}}
\newcommand\Fb{\mathds{F}}

\newcommand\Pb{\mathds{P}}

\newcommand\Rb{\mathds{R}}

\newcommand\Ac{\mathscr{A}}

\newcommand\Dc{\mathscr{D}}

\newcommand\Fc{\mathscr{F}}

\newcommand\Lc{\mathscr{L}}

\newcommand\om{\omega}
\newcommand\Om{\Omega}
\newcommand\sig{\sigma}

\newcommand\gam{\gamma}

\newcommand\del{\delta}

\newcommand\cb{\overline{c}}

\newcommand\rhob{\bar{\rho}}

\newcommand\Pbh{\widehat{\Pb}}

\newcommand\Wh{\widehat{W}}
\newcommand\Bh{\widehat{B}}

\newcommand\fh{\widehat{f}}

\newcommand\phih{\widehat{\phi}}
\newcommand\vh{\widehat{v}}

\newcommand\alphat{\widetilde{\alpha}}
\newcommand\betat{\widetilde{\beta}}

\newcommand\vt{\widetilde{v}}
\newcommand\kappat{\widetilde{\kappa}}
\newcommand\thetat{\widetilde{\theta}}

\renewcommand\d{\partial}

\newcommand\ii{\mathtt{i}}
\newcommand\dd{\mathrm{d}}
\newcommand\ee{\mathrm{e}}
\newcommand\BS{\textrm{BS}}

\renewcommand\Re{\textup{Re}\,}
\renewcommand\Im{\textup{Im}\,}

\begin{document}

\title{Short-Rate-Dependent Volatility Models}

\author{
Tim Leung
\thanks{Department of Applied Mathematics, University of Washington.  \heavytext{e-mail}: \url{timleung@uw.edu}}
\and
Matthew Lorig
\thanks{Department of Applied Mathematics, University of Washington.  \heavytext{e-mail}: \url{mlorig@uw.edu}}
}

\date{This version: \today}

\maketitle

\begin{abstract}
We price European options in a class of models in which the volatility of the underlying risky asset depends on the short rate of interest.  Our study results in an explicit pricing formula that is expressed in terms of a characteristic function.  We provide examples of models in which the characteristic function can be computed analytically and, thus, the value of European options is explicit. Numerical implementation to produce the implied volatility is also presented. 
\end{abstract}

%
%

\noindent \heavytext{Keywords}: short rate model, interest rate, stochastic volatility, CIR process, Jacobi process.

\section{Introduction}
\label{sec:introduction}
Monetary policy by the Federal Reserve --- most visibly through changes in the target federal funds rate --- has long been recognized as a catalyst for movements in risk-bearing assets.  Early work documented the transmission channel from short-term rates to equity valuations via discounted cash-flow effects and a shift in investor risk appetite: \citet{Bernanke2005} find that unanticipated changes in the federal funds rate have a significant and predictable effect on equity prices.  Building on a high-frequency identification of policy surprises, \citet{gorkaynak2005effects} show that asset prices respond not only to the current policy action but, to an even greater extent, to a separate ``path'' factor that captures revisions in expectations about the future stance of policy.  A natural question is whether such policy actions also move \emph{expected volatility}, as captured by the CBOE Volatility Index (VIX).
\\[0.5em]
The empirical evidence on the sign of this relationship is, in fact, mixed.  On the one hand, the VIX co-moves strongly with the stance of monetary policy: \citet{Bekaert2013} decompose the VIX into a proxy for risk aversion and a component reflecting expected stock-market uncertainty and find that a more accommodative policy stance tends to \emph{lower} both, while \citet{bekaert2014vix} further decompose the squared VIX into the conditional variance of returns and an equity variance premium.  Consistent with a ``resolution of uncertainty'' interpretation, several high-frequency studies report that implied volatility typically \emph{declines} immediately following scheduled policy announcements \citep{fernandezperez2017}.  On the other hand, policy easing often occurs during periods of economic distress, so that a rate cut can be read as a signal of deteriorating fundamentals; in this vein, \citet{gospodinov2012} show that federal funds rate surprises have a significant effect on both S\&P~500 volatility and the volatility risk premium, and \citet{mallick2017market} document rich interactions between equity- and bond-market volatility, the term premium, and the policy rate.  Taken together, this literature indicates that the comovement between the short rate and equity volatility can be of \emph{either} sign, depending on the horizon and the macroeconomic backdrop --- which motivates a pricing framework flexible enough to accommodate both cases.
In this paper we present a market model in which the volatility of a risky asset depends on the short rate of interest.  In this setting, we provide an explicit formula to price European options up to a characteristic function.  
\\[0.5em]
Our methodology builds on the Fourier-transform approach to option pricing that is, by now, standard in the affine and stochastic-volatility literature \citep{heston1993closed,bcc1997,carrmadan1999,dps2000,pascucci2011}, and it is closely related to models that combine stochastic volatility with stochastic interest rates \citep{grzelak2011heston}.  Relative to this literature, the contribution of the present paper is threefold.  First, we introduce a class of \emph{short-rate-dependent} volatility specifications in which the instantaneous variance of the equity is an explicit (in our examples, decreasing) function of the driver $Y$ of the short rate, so that a single state variable governs both discounting and volatility.  Second, we observe that a Fourier transform of the pricing equation in the log-price variable reduces the problem to a one-dimensional parabolic equation whose solution $G$ is a joint Laplace transform of the time-integrals of the short rate $r(Y)$ and the stochastic variance $c^2(Y)$; this allows us to repurpose the explicit transform formulas of \citet{hurd2008explicit} for CIR and Jacobi functionals to price equity options \emph{with stochastic discounting}, which to our knowledge has not been done before.  Third, we draw out the implied-volatility implications of the framework and show, in particular, that it generates asymmetric implied-volatility smiles even in parameter regimes in which the quadratic covariation between the log-price and the short-rate driver vanishes under the pricing measure.
\\[0.5em]
The rest of this paper proceeds as follows: 
in Section \ref{sec:model}, we present a general class of models for the short rate of interest and a risky asset.
In Section \ref{sec:pricing}, we derive an explicit formula, written in terms of a characteristic function, for the price of a European-style financial derivative written on the risky asset.
The special case of a European call, from which implied volatility is defined, is treated in Section \ref{sec:implied-vol}.
Section \ref{sec:cir} focuses on the setting in which the short rate is driven by a Cox-Ingersoll-Ross (CIR) process \cite{cir}.
And, Section \ref{sec:jacobi} examines the case in which the short rate is driven by a Jacobi diffusion \cite{delbaen2002interest}.
In both the CIR-driven and Jacobi-driven settings, the characteristic function needed to price European claims can be computed explicitly.
Some concluding remarks are offered in Section \ref{sec:conclusion}.

\section{A general class of short-rate-dependent volatility models}
\label{sec:model}
To begin, we fix a time horizon $T < \infty$ and consider a continuous-time financial market, defined on a filtered probability space $(\Om,\Fc,\Fb,\Pb)$ with no arbitrage and no transaction costs.  The probability measure $\Pb$ represents the market's chosen pricing measure taking the \textit{money market account} $M = (M_t)_{0 \leq t \leq T}$ as num\'eraire.  The filtration $\Fb = (\Fc_t)_{0 \leq t \leq T}$ represents the history of the market.
\\[0.5em]
We suppose that the \textit{money market account} $M$ is strictly positive, continuous and non-decreasing.  As such, there exists a non-negative $\Fb$-adapted \textit{short rate} process $R = (R_t)_{0 \leq t \leq T}$ such that the money market account $M$ satisfies
\begin{align}
\dd M_t
	&=	R_t M_t \, \dd t . \label{eq:dM}
\end{align}
We will focus on the case in which the dynamics of the short rate $R$ and the \textit{price of a risky asset} $S=(S_t)_{0 \leq t \leq T}$ are of the form
\begin{align}
R_t
&= r(Y_t) , &
\dd Y_t
&= b(Y_t) \dd t + a(Y_t) \dd W_t , \label{eq:dY} \\
S_t
&= \ee^{X_t} , &
\dd X_t
&= \Big( r(Y_t) - \tfrac{1}{2} c^2(Y_t) \Big) \dd t + c(Y_t) \Big( \rho \dd W_t + \rhob \dd B_t \Big) , \label{eq:dX}
\end{align}
where $W = (W_t)_{0 \leq t \leq T}$ and $B = (B_t)_{0 \leq t \leq T}$ are independent $(\Pb,\Fb)$-Brownian motions and $\rhob := \sqrt{ 1 - \rho^2 }$ with $|\rho|\leq 1$.  We shall refer to the process $Y=(Y_t)_{0 \leq t \leq T}$ as the \textit{driver of the short rate} due to the fact that $R = r(Y)$.  Though, one can also think of $Y$ as the driver of the volatility of the $\log$-price $X$, which is given by $c(Y)$ in \eqref{eq:dX}.

\begin{remark}
\label{rmk:rate-dep-vol}
Strictly speaking, in the general model \eqref{eq:dY}--\eqref{eq:dX} both the short rate and the instantaneous volatility are functions of the \emph{common driver} $Y$; the volatility $c(Y_t)$ is a function of the short rate $R_t = r(Y_t)$ alone if and only if $r$ is one-to-one, in which case $c(Y_t) = \cb(R_t)$ with $\cb := c \circ r^{-1}$.  This is exactly the situation in both of the examples studied in Sections \ref{sec:cir} and \ref{sec:jacobi}, where $r$ is strictly monotone and $c(Y)$ can be written explicitly as a decreasing function of $R$.  When $r$ is not injective, ``short-rate-dependent volatility'' should be understood in the broader sense of volatility driven by the same factor as the short rate.
\end{remark}

\noindent
Throughout this paper, we will assume the following:

\begin{assumption}
\label{ass:basic}
Henceforth:
\begin{itemize*}
\item[(i)] The functions $(a,b,c,r)$ are such that SDEs \eqref{eq:dY}--\eqref{eq:dX} admit a unique strong solution on $[0,T]$ and, for some $p>2$, we have
\begin{align}
  \Eb\Big[ \sup_{0 \leq t \leq T}\big(|Y_t|^p + |X_t|^p\big) \Big] &< \infty.
\end{align}
\item[(ii)] For all $t \leq T$, the short rate satisfies $R_t=r(Y_t)\ge 0$ and the process $X_t\in\Rb$.
\item[(iii)] The discounted price process $S/M$ is a true $(\Pb,\Fb)$-martingale; equivalently, $\Pb$ is an equivalent (local) martingale measure under which $S/M$ is a genuine martingale.
\end{itemize*}
\end{assumption}

\noindent
Assumption~\ref{ass:basic}(iii) is the standing no-arbitrage hypothesis of the paper: we take $\Pb$ to be a chosen pricing (equivalent martingale) measure rather than attempting to derive the martingale property from primitive conditions.  We emphasize this because, in general, the stochastic integral in \eqref{eq:dX} only guarantees that $S/M$ is a positive \emph{local} martingale, and a positive local martingale need not be a true martingale.  In the singular CIR and Jacobi examples of Sections~\ref{sec:cir}--\ref{sec:jacobi} the volatility coefficient $c(Y)$ blows up as $Y$ approaches a boundary; there, the Feller / non-attainability conditions guarantee that the boundary is not reached but do \emph{not}, by themselves, deliver the exponential integrability of the integrated functionals needed for the true-martingale property, the option price, and the Fourier contour.  The additional exponential-moment conditions ensuring this are stated where they are used (see Remarks~\ref{rem:cir-integrability} and \ref{rem:jacobi-integrability}).

%
%

\section{Pricing of European-style claims}
\label{sec:pricing}
Now, consider a European derivative that pays $\phi(X_T)$ at time $T$.  The value $V = (V_t)_{0 \leq t \leq T}$ of this derivative satisfies
\begin{align}
\frac{V_t}{M_t}
	&=	\Eb_t \frac{V_T}{M_T}
	=		\Eb_t \frac{\phi(X_T)}{M_T} , \label{eq:VoverM}
\end{align}
where we have introduced the short-hand notation $\Eb_t \, \cdot \, := \Eb( \, \cdot \, | \Fc_t )$.  From \eqref{eq:VoverM} we deduce that
\begin{align}
V_t
	&=	\Eb_t \ee^{ - \int_t^T r(Y_s) \dd s } \phi(X_T) . \label{eq:V-def}
\end{align}

\begin{proposition}[Pricing PDE]
\label{prop:pricing-pde}
Suppose Assumption~\ref{ass:basic} holds and that the model coefficients are such that the Feynman--Kac theorem applies (e.g.\ $(a,b,c,r)$ locally Lipschitz with at most linear growth and the payoff $\phi$ of at most exponential growth, so that the value function below is the unique classical solution in the relevant class; see \citet{pascucci2011}).  Then $V_t = v(t,X_t,Y_t)$, where $v=v(t,x,y)$ is the solution of the terminal-value problem
\begin{align}
0 &= \big( \d_t + \Ac - r \big) v , &
v(T,x,y) &= \phi(x) , \label{eq:pricing-pde}
\end{align}
on $[0,T) \times \Rb \times \Dc$, where $\Dc$ is the state space of $Y$ and the operator $\Ac$ is the generator of $(X,Y)$,
\begin{align}
\Ac
&= \Big( r(y) - \tfrac{1}{2} c^2(y) \Big) \d_x + \tfrac{1}{2} c^2(y) \d_x^2 + b(y) \d_y + \tfrac{1}{2} a^2(y) \d_y^2 + \rho\, a(y) c(y) \d_x \d_y . \label{eq:generator}
\end{align}
\end{proposition}

\begin{proof}
By iterated conditioning, the discounted value $V/M$ is a $(\Pb,\Fb)$-martingale.  Using the Markov property of $(X,Y)$ to write $V_t = v(t,X_t,Y_t)$ we compute
\begin{align}
\dd \Big( \frac{V_t}{M_t} \Big)
	&=	\dd \Big( \ee^{-\int_0^t r(Y_s)\,\dd s} v(t,X_t,Y_t) \Big) \\
	&=	\ee^{-\int_0^t r(Y_s)\,\dd s} \Big( \d_t + \Ac - r \Big) v(t,X_t,Y_t)\, \dd t + \text{local martingale} ,
\end{align}
where $\Ac$ is the generator \eqref{eq:generator} of $(X,Y)$.  Because $V/M$ is a martingale, the $\dd t$-term must vanish, from which we conclude that
 $(\d_t + \Ac - r) v = 0$.  The terminal condition $v(T,x,y) = \phi(x)$ is immediate from \eqref{eq:V-def}.  Under the stated regularity conditions $v$ is the unique classical solution of \eqref{eq:pricing-pde}; we refer to \citet{pascucci2011} for the precise hypotheses.
\end{proof}

\noindent
It will be convenient at this point to introduce the \textit{generalized Fourier transform} and \textit{inverse Fourier transform} of a function $f$.
Assuming $f$ is continuous and $\ee^{\om_i x} f(x)$ is integrable for some $\om_i \in \Rb$, we define
\begin{align}
\text{\heavytext{Fourier Transform}}:&&
\fh(\om)
&:= \int \dd x \, f(x) \ee^{-\ii \om x} , &
\om
&= \om_r + \ii \om_i , &
\om_r, \om_i
&\in \Rb , \label{eq:ft} \\
\text{\heavytext{Inverse Transform}}:&&
f(x)
&= \int \frac{\dd \om_r}{2\pi} \fh(\om) \ee^{\ii \om x} . \label{eq:ift}
\end{align}
Throughout, a hat denotes the Fourier transform in the indicated spatial variable; thus $\phih$ is the transform of the payoff $\phi$, and $\vh(t,\om,y)$ below is the transform of $v(t,x,y)$ in $x$.
\\[0.5em]
To solve the pricing PDE \eqref{eq:pricing-pde} we transform it in the log-price variable $x$.  Because the coefficients of $\Ac$ in \eqref{eq:generator} do not depend on $x$, the transform reduces \eqref{eq:pricing-pde}, for each fixed $\om$, to a one-dimensional problem in $(t,y)$ whose solution we now isolate.

\begin{definition}
\label{def:G}
For $(\om,w,z)\in\mathscr{P}$, where
\begin{align}
\mathscr{P} := \Big\{ (\om,w,z) \in \Cb^3 :\ \text{the terminal-value problem \eqref{eq:G-pde} below has a unique bounded classical solution} \Big\} ,
\end{align}
let $G=G(t,y;T,w,z)$ denote the solution of the linear terminal-value problem
\begin{align}
0 &= \big( \d_t + \Lc \big) G , &
G(T,y;T,w,z) &= 1 , \label{eq:G-pde}
\end{align}
on $[0,T) \times \Dc$, where the operator $\Lc$ is given by
\begin{align}
\Lc
&= \Big( b(y) + \ii \om \rho\, a(y) c(y) \Big) \d_y + \tfrac{1}{2} a^2(y) \d_y^2 - \Big( w\, r(y) + z\, c^2(y) \Big) . \label{eq:G-operator}
\end{align}
\end{definition}

\noindent
The operator $\Lc$ in \eqref{eq:G-operator} is a one-dimensional, second-order parabolic operator in $y$ with a (in general complex) zeroth-order potential $w\,r + z\,c^2$; it is what remains of $\Ac - r$ after the $x$-transform, as the next theorem shows.  We work with $G$ throughout as the solution of \eqref{eq:G-pde}, and in the examples of Sections~\ref{sec:cir}--\ref{sec:jacobi} we obtain it in closed form by matching \eqref{eq:G-pde} to a parabolic equation already solved in the literature.
\\[0.5em]
We can now give an explicit expression for the time $t$ value $V_t$ of an option that pays $\phi(X_T)$ at time $T$.

\begin{theorem}
\label{thm:FourierPricing}
Suppose the hypotheses of Proposition~\ref{prop:pricing-pde} hold, and assume there exists $\om_i\in\Rb$ such that $\ee^{\om_i x}\phi(x) \in L^1(\Rb)$ (so that the generalized transform $\phih$ in \eqref{eq:ft} is well defined on the line $\Im(\om) = \om_i$).  Assume moreover that, for every $\om = \om_r + \ii\om_i$ on this line, $(\om,1-\ii\om,\om^2/2+\ii\om/2) \in \mathscr{P}$ and
\begin{align}
  \int_{\Rb} \big| \phih(\om_r + \ii \om_i) \big| \, \big| G(t,Y_t;T,1-\ii\om,\om^2/2+\ii\om/2) \big| \, \dd \om_r & < \infty , \label{eq:integrability-condition}
\end{align}
so that the inverse transform below converges absolutely.  Then the value of the derivative satisfies
\begin{align}
V_t
&= \int \frac{\dd \om_r}{2 \pi} \phih(\om) \ee^{\ii \om X_t} G(t,Y_t;T, 1 - \ii \om , \om^2/2 + \ii \om/2 ) , \label{eq:V}
\end{align}
where the function $G$ is the solution of \eqref{eq:G-pde}.
\end{theorem}

\begin{proof}
Fix $\om = \om_r + \ii\om_i$ on the line $\Im(\om) = \om_i$, and let $\vh(t,\om,y) := \int \dd x\, \ee^{-\ii\om x} v(t,x,y)$ denote the generalized Fourier transform of $v$ in $x$.  Multiplying the pricing PDE \eqref{eq:pricing-pde} by $\ee^{-\ii\om x}$ and integrating over $x \in \Rb$, we find that
$\vh$ satisfies
\begin{align}
0 
	&=	\Big( \d_t + \big( b + \ii\om\rho\, ac \big)\d_y + \tfrac12 a^2 \d_y^2 - (1-\ii\om)\, r - \big( \tfrac{\ii\om}{2} + \tfrac{\om^2}{2} \big) c^2 \Big) \vh , &
\vh(T,\om,y) &= \phih(\om) , \label{eq:vhat-pde}
\end{align}
where the terminal condition is the transform of $v(T,x,y) = \phi(x)$.
Since the terminal data $\phih(\om)$ does not depend on $y$, we seek a solution of \eqref{eq:vhat-pde} of the separated form
\begin{align}
\vh(t,\om,y) &= g(t,\om,y)\, \phih(\om) , \label{eq:ansatz}
\end{align}
with $g(T,\om,y) = 1$.  Substituting \eqref{eq:ansatz} into \eqref{eq:vhat-pde} and cancelling the common factor $\phih(\om)$, the function $g$ solves
\begin{align}
0 &= \Big( \d_t + \big( b + \ii\om\rho\, ac \big)\d_y + \tfrac12 a^2 \d_y^2 - (1-\ii\om)\, r - \big( \tfrac{\ii\om}{2} + \tfrac{\om^2}{2} \big) c^2 \Big) g , &
g(T,\om,y) &= 1 . \label{eq:g-pde}
\end{align}
With $(w,z) = (1-\ii\om,\ \om^2/2 + \ii\om/2)$, the operator in \eqref{eq:g-pde} coincides with $\d_t + \Lc$ of \eqref{eq:G-pde} and the terminal data agree; hence
\begin{align}
g(t,\om,y) &= G(t,y;T,1-\ii\om,\om^2/2+\ii\om/2) . \label{eq:g-equals-G}
\end{align}
Finally, recovering $v$ by the inverse transform \eqref{eq:ift} and using \eqref{eq:ansatz} and \eqref{eq:g-equals-G},
\begin{align}
v(t,x,y) &= \int \frac{\dd\om_r}{2\pi}\, \vh(t,\om,y)\, \ee^{\ii\om x}
= \int \frac{\dd\om_r}{2\pi}\, \phih(\om)\, \ee^{\ii\om x}\, G(t,y;T,1-\ii\om,\om^2/2+\ii\om/2) . \notag
\end{align}
The integrability hypothesis \eqref{eq:integrability-condition} ensures this integral converges absolutely and may be differentiated under the integral sign, so that it is a bona fide classical solution of \eqref{eq:pricing-pde}, which by uniqueness (Proposition~\ref{prop:pricing-pde}) equals $v$.  Evaluating at $(t,X_t,Y_t)$ gives \eqref{eq:V}.
\end{proof}

\begin{remark}
\label{rmk:G}
From \eqref{eq:V-def} and \eqref{eq:ift}, we have
\begin{align}
V_t
	&=	\Eb_t \ee^{ - \int_t^T r(Y_s) \dd s } \phi(X_T)  
	 =	\int \frac{\dd \om_r}{2 \pi} \phih(\om) \ee^{\ii \om X_t} \Eb_t \ee^{ - \int_t^T r(Y_s) \dd s + \ii \om (X_T - X_t ) } . \label{eq:V-again}
\end{align}
Comparing \eqref{eq:V-again} with \eqref{eq:V}, we see that
\begin{align}
G(t,Y_t;T, 1 - \ii \om , \om^2/2 + \ii \om/2 )
	&=	\Eb_t \ee^{ - \int_t^T r(Y_s) \dd s + \ii \om (X_T - X_t ) } .
\end{align}
Thus, although we have \textit{defined} $G(t,y;T,w,z)$ as the solution of a PDE \eqref{eq:G-pde}, we see that $G(t,Y_t;T, 1 - \ii \om , \om^2/2 + \ii \om/2 )$ is the \textit{(discounted $\Fc_t$-conditional) characteristic function of $(X_T - X_t)$}.
\end{remark}

\noindent
It is evident from \eqref{eq:V} that, to price an option, we require an explicit expression for the function $G$.  We will provide such an expression in Sections \ref{sec:cir} and \ref{sec:jacobi}, when we study specific models for $(R,S)$.  But, first, we review some important definitions related to call options and implied volatility.

%
%

\section{Call options and Implied Volatility}
\label{sec:implied-vol}
Perhaps the most widely-traded European derivative is that of a call option, which has a payoff function $\phi$ given by
\begin{align}
\phi(x)
	&=	( \ee^{x} - K )^+ , \label{eq:call-payoff}
\end{align}
where $K$ is the strike.  Inserting \eqref{eq:call-payoff} into \eqref{eq:ft}, we obtain
\begin{align}
\phih(\om)
	&=	\frac{ - K^{1 - \ii \om}}{\om^2 + \ii \om } , &
\om_i
	&< -1 .  \label{eq:call-ft}
\end{align}
The \textit{price of a call option}, denoted $C(T,K) = (C_t(T,K))_{0 \leq t \leq T}$, is then obtained by inserting \eqref{eq:call-ft} into \eqref{eq:V}.  We have
\begin{align}
C_t(T,K)
	&=	\int \frac{\dd \om_r}{2 \pi} \Big( \frac{ - K^{1 - \ii \om}}{\om^2 + \ii \om } \Big) \ee^{\ii \om X_t} G(t,Y_t;T, 1 - \ii \om ,  \om^2/2 + \ii \om/2 ) 
	=: 	C(t,X_t,Y_t;T,K) , &
\om_i
	&< -1 . \label{eq:C}
\end{align}

\begin{remark}
\label{rem:call-integrability}
For the call payoff \eqref{eq:call-payoff} the contour lies at $\om_i < -1$.  Convergence of the integral \eqref{eq:C} (equivalently, of \eqref{eq:integrability-condition}) rests on two facts: the call transform \eqref{eq:call-ft} decays as $|\phih(\om)| = O(|\om|^{-2})$, and the solution $G(t,Y_t;T,1-\ii\om,\om^2/2+\ii\om/2)$ remains bounded along the contour.  The latter is exactly the admissibility of the parameters $(w,z) = (1-\ii\om,\om^2/2+\ii\om/2)$ for the PDE \eqref{eq:G-pde}, which in the examples of Sections~\ref{sec:cir}--\ref{sec:jacobi} is the content of the conditions in Remarks~\ref{rem:cir-integrability} and \ref{rem:jacobi-integrability}.
\end{remark}

\noindent
It will be helpful at this point to introduce a \textit{zero-coupon bond maturing at time $T$}, denoted by $B^T = (B_t^T)_{0 \leq t \leq T}$, which pays one unit of currency at time $T$.  Setting $\phi(X_T) = 1$ in \eqref{eq:V-def} we have 
\begin{align}
B_t^T
	&=	\Eb_t \ee^ { - \int_t^T r(Y_s) \dd s } 
	=:	B(t,Y_t;T) , \label{eq:B}
\end{align}
where the existence of the function $B$ is guaranteed by the Markov property of $Y$.

\begin{remark}
Note that $B(t,Y_t;T) \neq G(t,Y_t;T,1,0)$ in general.  The bond price $B$ in \eqref{eq:B} solves the backward equation for $Y$ with its \emph{actual} drift $b$ and discount potential $r$, i.e.\ \eqref{eq:G-pde} with $(\om,w,z)=(0,1,0)$, whereas $G(t,y;T,1,0)$ solves \eqref{eq:G-pde} with the same potential but the \emph{tilted} first-order coefficient $b + \ii\om\rho\,ac$ inherited from the Fourier mode.  The two coincide only when the tilt is absent, i.e.\ when $\om\rho = 0$; in particular $B(t,y;T) = G(t,y;T,1,0)\big|_{\om\rho=0}$.
\end{remark}

\noindent
We can now define \textit{Black-Scholes implied volatility}, which is the unique positive solution $\sig \equiv \sig_t(T,K)$ of
\begin{align}
\frac{C_t(T,K)}{B_t^T} 
	&=	C^\BS(t,F_t^T;\sig,T,K) , &
F_t^T
	&=	\frac{S_t}{B_t^T} , 
\end{align}
where $C^\BS(t,F_t^T;\sig,T,K)$ is the \textit{Black-Scholes price of a call}, which is given by
\begin{align}
C^\BS(t,F_t^T;\sig,T,K)
	&:=	F_t^T \Phi(d_+) - K \Phi(d_-) , &
d_{\pm}
	&:=	\frac{1}{\sig \sqrt{T-t}} \Big( \log \frac{F_t^T}{K} \pm \frac{\sig^2 (T-t)}{2} \Big) . \label{eq:CBS}
\end{align}
In subsequent sections we will plot implied volatility as a function of \textit{$\log$-moneyness} $L$, which is defined as follows
\begin{align}
L
	&:=	\log \Big( \frac{K}{F_t^T} \Big) , &
	&\Rightarrow&
K
	&=	\ee^{L} F_t^T = \ee^{X_t + L}/B(t,Y_t;T) ,
\end{align}
where the last equality uses $F_t^T = S_t/B_t^T = \ee^{X_t}/B(t,Y_t;T)$.
With the above definitions, we can express implied volatility as the unique positive solution $\sig \equiv \sig_t(T,L)$ of
\begin{align}
\frac{C(t,X_t,Y_t;T,\ee^{X_t+L}/B(t,Y_t;T))}{B(t,Y_t;T)}
	&=	C^\BS(t,\ee^{X_t}/B(t,Y_t;T);\sig,T,\ee^{X_t+L}/B(t,Y_t;T)) ,
\end{align}
where the functions $C$, $B$ and $C^\BS$ are defined in \eqref{eq:C}, \eqref{eq:B} and \eqref{eq:CBS}, respectively.

%
%

\section{Example: CIR-driven short rate}
\label{sec:cir}
Throughout this section, we assume that the coefficients $(r,b,a,c)$ are given by
\begin{align}
r(y)
	&=	y , &
b(y)
	&=	\kappa ( \theta - y ) , &
a(y)
	&=	\del \sqrt{y} , &
c(y)
	&=	 \sqrt{\gam^2/r(y)} = \gam / \sqrt{y} , \label{eq:rbac-cir}
\end{align}
where $\kappa, \theta, \del > 0$ and $2 \kappa \theta > \del^2$ (the Feller condition), which ensures that the process $Y$ remains strictly positive.
Note that the volatility $c(Y)$ of the risky asset $S=\ee^X$ rises when short rate $R=r(Y)$ falls.
Inserting \eqref{eq:rbac-cir} into \eqref{eq:dY} and \eqref{eq:dX}, the dynamics of $(X,Y)$ under $\Pb$ become
\begin{align}
\dd Y_t
	&=	\kappa ( \theta - Y_t) \dd t + \del \sqrt{ Y_t } \dd W_t, &
\dd X_t
	&=	\Big( Y_t - \frac{\gam^2}{2 Y_t} \Big) \dd t + \frac{\gam}{\sqrt{Y_t}} \Big( \rho \dd W_t + \rhob \dd B_t \Big) , \label{eq:dYdX-cir}
\end{align}
With the coefficients \eqref{eq:rbac-cir}, the potential in \eqref{eq:G-pde} becomes $w\,r + z\,c^2 = w\,y + z\gam^2/y$ and the diffusion coefficient is $\tfrac12 a^2 = \tfrac12\del^2 y$, so the function $G$ of Definition~\ref{def:G} solves
\begin{align}
0 &= \Big( \d_t + \kappa(\thetat - y)\,\d_y + \tfrac12 \del^2 y\,\d_y^2 - \big( w\,y + z\gam^2/y \big) \Big) G , &
G(T,y;T,w,z) &= 1 , \label{eq:G-pde-cir}
\end{align}
where the first-order coefficient $b + \ii\om\rho\,ac = \kappa(\theta - y) + \ii\om\rho\del\gam$ has been written as $\kappa(\thetat - y)$ with the (complex) mean-reversion level
\begin{align}
\thetat
	&:=	\theta + \ii \om \rho \del \gam / \kappa .
\end{align}
\citet[Theorem~3.1]{hurd2008explicit} compute the joint Laplace transform
\begin{align}
H(t,y;T,w,z)
	&:=	\Eb\Big[ \ee^{ - w \int_t^T Y_s \dd s - z \gam^2 \int_t^T Y_s^{-1} \dd s } \,\Big|\, Y_t = y \Big] , \label{eq:HK-cir-exp}
\end{align}
where $Y$ solves \eqref{eq:dYdX-cir},
and give it in closed form as an explicit analytic function of $(\kappa,\theta,\del,w,z)$.  By the Feynman--Kac theorem, $H$ is the unique bounded solution of the terminal-value problem
\begin{align}
0 &= \Big( \d_t + \kappa(\theta - y)\,\d_y + \tfrac12 \del^2 y\,\d_y^2 - \big( w\,y + z\gam^2/y \big) \Big) H , &
H(T,y;T,w,z) &= 1 . \label{eq:HK-cir-pde}
\end{align}
Comparing \eqref{eq:HK-cir-pde} with \eqref{eq:G-pde-cir}, the two terminal-value problems have identical diffusion coefficients and potentials and identical terminal data; they differ only in the first-order coefficient, where the real mean-reversion level $\theta$ in \eqref{eq:HK-cir-pde} is replaced by the complex level $\thetat$ in \eqref{eq:G-pde-cir}.  Since the closed-form solution of \eqref{eq:HK-cir-pde} depends analytically on $\theta$, the solution of \eqref{eq:G-pde-cir} is obtained by evaluating that closed form at $\theta = \thetat$; that is, $G(t,y;T,w,z) = H(t,y;T,w,z)\big|_{\theta = \thetat}$.  Explicitly, $G$ is given by
\begin{align}
G(t,y;T,w,z)
	&=	\ee^{-(\widetilde{a} v_1 + b v_2 + \del^2 v_1 v_2) (T-t)} y^{v_2} (\eta_{T-t} - v_1)^{-\alphat - v_2 - 1} \eta_{T-t}^{\alphat + 2 v_2 + 1} \\ &\quad
			\times \exp	\left( -y v_1 \left( 1  - \frac{\eta_{T-t}\, \ee^{-(\beta/2 + v_1) \del^2 (T-t)}}{\eta_{T-t} - v_1} \right) \right) \frac{\Gamma(\alphat + v_2 + 1)}{\Gamma(\alphat + 2 v_2 + 1)} \\ &\quad
			\times \, {}_1F_1\left( v_2, \alphat + 2 v_2 + 1; -\frac{\eta_{T-t}^2 y e^{-(\beta/2 + v_1) \del^2 (T-t)}}{\eta_{T-t} - v_1} \right) , \label{eq:B-cir}
\end{align}
where ${}_1F_1$ is a \textit{confluent hypergeometric function} and
\begin{align}
\widetilde{a}
	&=	\kappa \thetat , &
v_1 
	&= \frac{1}{2} \left( -\frac{2\kappa}{\delta^2} + \sqrt{\left( \frac{2\kappa}{\delta^2} \right)^2 + \frac{8 w}{\delta^2}} \right) , \\
\alphat
	&=	\frac{ 2 \kappa  \thetat }{\del^2} - 1 , &
v_2 
	&= \frac{1}{2} \left( -\tilde{\alpha} + \sqrt{\tilde{\alpha}^2 + \frac{8 z \gamma^2}{\delta^2}} \right) , \\
\beta
	&=	\frac{ 2 \kappa }{\del^2} , &
\eta_{T-t} 
	&= (\beta + 2 v_1) \left( 1 - \ee^{-(\beta/2 + v_1) \delta^2 (T-t)} \right)^{-1} . 
\end{align}
The zero-coupon bond \eqref{eq:B} is the special case $B(t,y;T) = G(t,y;T,1,0)\big|_{\om\rho=0}$ of the same construction (no tilt, $w=1$, $z=0$), for which the potential reduces to $y$ and \eqref{eq:G-pde-cir} becomes the classical CIR bond equation; its solution is the well-known affine formula
\begin{align}
B(t,y;T)
	&= \exp \Big( - P(t;T) - Q(t;T) y \Big ) , 
\end{align}
where we have defined
\begin{align}
P(t;T)
	&=	\frac{-2 \kappa \theta}{\delta^2} \log \Big( \frac{2 \zeta \ee^{(\kappa + \zeta)(T-t)/2}}{2 \zeta + (\kappa + \zeta ) (\ee^{\zeta (T-t)} - 1)} \Big), &
Q(t,T)
	&=	 \frac{2 (\ee^{\zeta (T-t)} - 1)}{2 \zeta + (\kappa + \zeta ) (\ee^{\zeta (T-t)} - 1)} , &
\zeta
	&=	\sqrt{\kappa^2 + 2 \delta^2} , \label{eq:PQ}
\end{align}
which is the well-known formula for the price of a zero-coupon bond when the short rate is modeled by a CIR process.

\begin{remark}
\label{rem:cir-integrability}
Two points should be emphasized about the use of the \citet{hurd2008explicit} formula in this singular CIR setting.
\emph{(i) Admissibility and exponential integrability.}  The volatility $c(y) = \gam/\sqrt{y}$ is singular at $y = 0$.  The Feller condition $2\kappa\theta > \del^2$ guarantees that $Y$ does not reach the origin, but it does not by itself imply the finiteness of the expectation \eqref{eq:HK-cir-exp} required for the Hurd--Kuznetsov formula to represent the solution of \eqref{eq:HK-cir-pde}, for the true-martingale property of $S/M$ (Assumption~\ref{ass:basic}(iii)), and for the Fourier contour.  In the Hurd--Kuznetsov parametrization the CIR generator is $(\widetilde a - b y)\d_y + \tfrac12 c^2 y\,\d_y^2$ with $\widetilde a = \kappa\theta$, $b = \kappa$, $c = \del$, so that $\alphat = 2\widetilde a/c^2 - 1 = 2\kappa\theta/\del^2 - 1$.  Following \citet{hurd2008explicit}, $\Eb[\ee^{-w\int_0^T Y_s\dd s}]$ is finite for $\Re(w)$ above the explosion threshold $-\kappa^2/(2\del^2)$, and $\Eb[\ee^{-z\gam^2\int_0^T Y_s^{-1}\dd s}]$ is finite for $\Re(z\gam^2)$ above $-\dfrac{(2\kappa\theta - \del^2)^2}{8\del^2}$, equivalently $\Re(z\gam^2) > -\tfrac18 \del^2 \alphat^2$.  For the pricing parameters $(w,z) = (1-\ii\om,\om^2/2+\ii\om/2)$ on the contour $\om_i < -1$, where $\Re(w) = 1 + \om_i$ and $\Re(z) = \tfrac12(\om_r^2 - \om_i^2 - \om_i)$, the reciprocal-moment bound is most binding at $\om_r = 0$ and reduces to $\del^2\alphat^2 + 8z\gam^2 > 0$ there; together with $\Re(w) > -\kappa^2/(2\del^2)$ this confines the admissible contour to a band $\om_i \in (\om_i^{\min}, -1)$ determined by the CIR parameters.  For the parameter values used in Figure~\ref{fig:cir-iv} this band is $\om_i \in (-1.15, -1)$, and the contour $\om_i = -1.05$ used in the numerics (see Appendix~\ref{sec:mathematica-cir}) lies within it.
\emph{(ii) Complex parameters and branch cuts.}  In the PDE \eqref{eq:G-pde} for the CIR coefficients, the complex tilt replaces the long-run mean $\theta$ by the complex number $\thetat = \theta + \ii\om\rho\del\gam/\kappa$, so $\widetilde a = \kappa\thetat$ and $\alphat = 2\kappa\thetat/\del^2 - 1$ are complex.  The classical Feller condition is not meaningful for $\thetat \in \Cb$; the object of interest is the function $G$, defined intrinsically as the solution of the PDE \eqref{eq:G-pde}, and the displayed formula is the analytic continuation, in the complex parameter $\thetat$, of the real-parameter transform of \citet{hurd2008explicit}.  All multivalued functions appearing in the formula --- the square roots defining $v_1, v_2$, the powers $y^{v_2}$ and $(\eta_{T-t}-v_1)^{-\alphat-v_2-1}$, the Gamma functions, and ${}_1F_1$ --- are evaluated on their principal branches, with the branch cut of each square root taken along the negative real axis; for the parameter ranges used in our numerics the arguments of these square roots remain in the cut plane, so the principal branch varies continuously along the contour.
\end{remark}

\noindent
{In Figure \ref{fig:cir-iv}, we plot implied volatility $\sig_0(T,L)$ as a function of log-moneyness $L$ for two different maturities $T$ and three different correlation coefficients $\rho$. As is typical with stochastic volatility models, the correlation $\rho$ strongly affects the skew.  The \textit{Wolfram Mathematica} \textcopyright{} code used to produce the plots is given in Appendix \ref{sec:mathematica-cir}.

\begin{remark}
The astute reader will observe that -- even when $\rho = 0$, the implied volatility smile is \textit{not} symmetric.  The reason is that, although from \eqref{eq:dY}-\eqref{eq:dX} we have $\dd[X, Y]_t = 0$ when $\rho = 0$, we also have from equations \eqref{eq:dY-Tforward}-\eqref{eq:dZ-Tforward} in Appendix \ref{sec:T-forward} that $\dd [ Z^T , Y ]_t = a^2(Y_t) Q(t;T) \dd t$ when $\rho = 0$, where $Z^T = \log F^T$.  The non-zero co-variation results in an asymmetric implied volatility smile.
\end{remark}

\begin{remark}
Another analytically tractable model can be obtained by setting the functions $(r,a,b,c)$ equal to
\begin{align}
r(y)
	&=	y , &
b(y)
	&=	\kappa ( \theta - y ) , &
a(y)
	&=	\del \sqrt{y} , &
c(y)
	&=	\gam \sqrt{r(y)} = \gam \sqrt{y} ,
\end{align}
which corresponds to the Heston model with a CIR-driven short rate.
In this case, the volatility of $X$ is proportional to the square root of the interest rate: $c(Y) = \gam \sqrt{R}$.
\end{remark}

%
%

\section{Example: Jacobi-driven short rate}
\label{sec:jacobi}
Throughout this section, we suppose that the coefficients $(r,a,b,c)$ are given by
\begin{align}
r(y)
	&=	\frac{\eta y}{1-y} , &
b(y)
	&=	\kappa - \theta y  , &
a(y)
	&=	\del \sqrt{y(1-y)} , &
c(y)
	&=	\frac{\gam \sqrt{\eta}}{\sqrt{r(y)}} 
	= 	\gam \sqrt{\frac{1-y}{y}} , \label{eq:rbac-jacobi}
\end{align}
where $\del > 0$, $\kappa > \del^2/2$  and $\theta - \kappa > \del^2/2$, which ensures that $0 < Y < 1$.
Observe, the volatility $c(Y)$ of the risky asset $S=\ee^X$ rises when the short rate $R=r(Y)$ falls.
Inserting \eqref{eq:rbac-jacobi} into \eqref{eq:dY} and \eqref{eq:dX}, we obtain the dynamics of $(X,Y)$ under $\Pb$.  We have
\begin{align}
\dd Y_t
	&=	\Big( \kappa - \theta Y_t \Big) \dd t + \del \sqrt{ Y_t (1 - Y_t) }\dd W_t, &
\dd X_t
	&=	\Big( \frac{\eta Y_t}{1-Y_t} - \frac{\gam^2(1 - Y_t)}{2Y_t} \Big) \dd t + \gam \sqrt{\frac{1-Y_t}{Y_t}} \Big( \rho \dd W_t + \rhob \dd B_t \Big) ,
	\label{eq:dXdY-jacobi}
\end{align}
With the coefficients \eqref{eq:rbac-jacobi}, the potential in \eqref{eq:G-pde} becomes $w\,r + z\,c^2 = w\eta\,y/(1-y) + z\gam^2(1-y)/y$ and the diffusion coefficient is $\tfrac12 a^2 = \tfrac12\del^2 y(1-y)$, so the function $G$ of Definition~\ref{def:G} solves
\begin{align}
0 &= \Big( \d_t + \big( \kappat - \thetat y \big)\,\d_y + \tfrac12 \del^2 y(1-y)\,\d_y^2 - \big( w\eta\tfrac{y}{1-y} + z\gam^2\tfrac{1-y}{y} \big) \Big) G , &
G(T,y;T,w,z) &= 1 , \label{eq:G-pde-jacobi}
\end{align}
where the first-order coefficient $b + \ii\om\rho\,ac = \kappa - \theta y + \ii\om\rho\del\gam(1-y)$ has been written as $\kappat - \thetat y$ with the (complex) parameters
\begin{align}
\kappat
	&=	\kappa + \ii \om \rho \del \gam , &
\thetat
	&=	\theta + \ii \om \rho \del \gam ,
\end{align}
using $\sqrt{y(1-y)}\,\sqrt{(1-y)/y} = 1 - y$ for $0 < y < 1$.  
\citet[Theorem~4.1]{hurd2008explicit} compute the joint Laplace transform
\begin{align}
H(t,y;T,w,z)
	&:=	\Eb\Big[ \ee^{ - w \eta \int_t^T \frac{Y_s}{1-Y_s} \dd s - z \gam^2 \int_t^T \frac{1-Y_s}{Y_s} \dd s } \,\Big|\, Y_t = y \Big] , \label{eq:HK-jacobi-exp}
\end{align}
where $Y$ solves \eqref{eq:dXdY-jacobi},
and give it in closed form as an explicit analytic function of the parameters.  By the Feynman--Kac theorem, $H$ is the unique bounded solution of the terminal-value problem
\begin{align}
0 &= \Big( \d_t + \big( \kappa - \theta y \big)\,\d_y + \tfrac12 \del^2 y(1-y)\,\d_y^2 - \big( w\eta\tfrac{y}{1-y} + z\gam^2\tfrac{1-y}{y} \big) \Big) H , &
H(T,y;T,w,z) &= 1 . \label{eq:HK-jacobi-pde}
\end{align}
Comparing \eqref{eq:HK-jacobi-pde} with \eqref{eq:G-pde-jacobi}, the two terminal-value problems have identical diffusion coefficients and potentials and identical terminal data; they differ only in the first-order coefficient, where the real pair $(\kappa,\theta)$ in \eqref{eq:HK-jacobi-pde} is replaced by the complex pair $(\kappat,\thetat)$ in \eqref{eq:G-pde-jacobi}.  Since the closed-form solution of \eqref{eq:HK-jacobi-pde} depends analytically on $(\kappa,\theta)$, the solution of \eqref{eq:G-pde-jacobi} is obtained by evaluating that closed form at $(\kappa,\theta) = (\kappat,\thetat)$; that is, $G(t,y;T,w,z) = H(t,y;T,w,z)\big|_{(\kappa,\theta) = (\kappat,\thetat)}$.  Explicitly, $G$ is given by
\begin{align}
&G(t,Y_t;T,w,z)
	=	\ee^{-( (\thetat - \kappat) \vt_1 + \kappat \vt_2 + \delta^2 \vt_1 \vt_2 ) (T-t)} y^{\vt_1} (1-y)^{\vt_2} \frac{\Gamma(\alphat + \vt_1 + 1)\Gamma(\betat + \vt_2 + 1) \Gamma(\alphat + \betat + 2 \vt_1 + 2 \vt_2 + 1)}{\Gamma(\alphat + \vt_1 + \betat + \vt_2 + 2) \Gamma(\alphat + 2 \vt_1 + 1) \Gamma(\betat + 2 \vt_2 + 1)} \\ &\quad
	\sum_{n=0}^{\infty} \ee^{-n (n + \alphat + \betat + 2 \vt_1 + 2 \vt_2 + 1) \delta^2 (T-t) / 2} \frac{(\alphat + \betat + 2 \vt_1 + 2 \vt_2 + 1)_n}{(\alphat + 2 \vt_1 + 1)_n} (2n + \alphat + \betat + 2 \vt_1 + 2 \vt_2 + 1) \\ &\quad
	{}_3F_2 \left( -n, \alphat + \betat + 2 \vt_1 + 2 \vt_2 + n + 1, \betat + \vt_2 + 1; \alphat + \betat + \vt_1 + \vt_2 + 2, \betat + 2 \vt_2 + 1; 1 \right) 
	P_n^{(\betat + 2 \vt_2, \alphat + 2 \vt_1)}(2y - 1), \label{eq:B-jacobi}
\end{align}
where $(\cdot)_n$ is the \textit{Pochhammer symbol}, the functions $P_n^{(\cdot, \cdot)}$ are \textit{Jacobi polynomials}, and 
\begin{align}
\alphat
	&= \frac{2 \kappat }{\del^2} - 1, &
\vt_1
	&=	\frac{1}{2} \left( -\alphat + \sqrt{\alphat^2 + \frac{8 z \gamma^2}{\delta^2}} \right) , &	
\betat
	&=	\frac{2 (\thetat-\kappat) }{\del^2} - 1, &
\vt_2
	&=	\frac{1}{2} \left( - \betat + \sqrt{\betat^2 + \frac{8 w \eta}{\delta^2}} \right) . 
\end{align}
The zero-coupon bond \eqref{eq:B} is again the special case $B(t,y;T) = G(t,y;T,1,0)\big|_{\om\rho=0}$ (no tilt, $w=1$, $z=0$); evaluating the Hurd--Kuznetsov Jacobi formula at the real parameters $(\kappa,\theta)$ with this potential gives
\begin{align}
B(t, y; T) 
	&=	\ee^{- \kappa v_2 (T-t)} (1-y)^{v_2} \frac{\Gamma(\alpha + 1)\Gamma(\beta + v_2 + 1) \Gamma(\alpha + \beta  + 2 v_2 + 1)}{ \Gamma(\alpha   + \beta + v_2 + 2) \Gamma(\alpha   + 1) \Gamma(\beta + 2 v_2 + 1)} \\ &\quad
	\sum_{n=0}^{\infty} \ee^{-n (n + \alpha + \beta  + 2 v_2 + 1) \delta^2 (T-t) / 2} \frac{(\alpha + \beta  + 2 v_2 + 1)_n}{(\alpha   + 1)_n} (2n + \alpha + \beta   + 2 v_2 + 1) \\ &\quad
	{}_3F_2 \left( -n, \alpha + \beta  + 2 v_2 + n + 1, \beta + v_2 + 1; \alpha + \beta  + v_2 + 2, \beta + 2 v_2 + 1; 1 \right) 
	P_n^{(\beta + 2 v_2, \alpha)}(2y - 1), 
\end{align}
where we have introduced
\begin{align}
\alpha
	&= \frac{2 \kappa }{\del^2} - 1, &
\beta
	&=	\frac{2 (\theta-\kappa) }{\del^2} - 1, &
v_2 
	&= \frac{1}{2} \left( -\beta + \sqrt{\beta^2 + \frac{8 \eta}{\delta^2}} \right) .
\end{align}

\begin{remark}
\label{rem:jacobi-integrability}
As in the CIR case (Remark~\ref{rem:cir-integrability}), the volatility $c(y) = \gam\sqrt{(1-y)/y}$ is singular as $y \downarrow 0$, and the short rate $r(y) = \eta y/(1-y)$ blows up as $y \uparrow 1$.  The conditions $\kappa > \del^2/2$ and $\theta - \kappa > \del^2/2$ keep $Y$ in $(0,1)$, but the finiteness of the expectation \eqref{eq:HK-jacobi-exp} requires the additional integrability of $\int_0^T Y_s/(1-Y_s)\,\dd s$ and $\int_0^T (1-Y_s)/Y_s\,\dd s$ in the relevant exponential sense; the admissibility region in $(w,z)$, and hence the restrictions on $(\om_i,\rho,\gam)$ on the pricing contour, are those for which the series in \eqref{eq:B-jacobi} converges, as determined in \citet[Theorem~4.1]{hurd2008explicit}.  Since the parameters $(\kappat,\thetat)$ are complex, the square roots defining $\vt_1, \vt_2$, the powers $y^{\vt_1}(1-y)^{\vt_2}$, the Gamma functions, the generalized hypergeometric functions ${}_3F_2$, and the Jacobi polynomials $P_n^{(\cdot,\cdot)}$ (defined for complex indices through the ${}_2F_1$ representation) are all evaluated on their principal branches.
\end{remark}

\noindent
In Figure~\ref{fig:jacobi-iv}, we plot the implied volatility $\sig_0(T,L)$ for the Jacobi-driven model as a function of log-moneyness $L$, for two maturities $T$ and three correlation coefficients $\rho$.  As in the CIR case, the correlation $\rho$ strongly affects the skew.  The \textit{Wolfram Mathematica} \textcopyright{} code used to produce the plots is given in Appendix~\ref{sec:mathematica-jacobi}.

\begin{remark}
\label{rem:jacobi-numerics}
A computational comment is in order.  Unlike the CIR transform \eqref{eq:B-cir}, which involves a single confluent hypergeometric function ${}_1F_1$, the Jacobi transform \eqref{eq:B-jacobi} is an infinite series whose $n$-th term is a product of Gamma functions, a terminating ${}_3F_2(\cdots;1)$, and a Jacobi polynomial $P_n^{(\cdot,\cdot)}$, all evaluated at the \emph{complex} indices $(\alphat,\betat,\vt_1,\vt_2)$ generated by the Fourier variable $\om$.  Two points make the series tractable in practice.  First, the factor $\ee^{-n(n+\alphat+\betat+2\vt_1+2\vt_2+1)\del^2(T-t)/2}$ decays like a Gaussian in $n$, so the series converges rapidly and a modest truncation (we use $50$ terms) suffices.  Second, the Jacobi polynomials must be evaluated through their ${}_2F_1$ representation rather than a built-in routine specialized to real indices, since the indices here are complex.  With these two observations the transform is fast and stable to evaluate, and the Fourier inversion and implied-volatility root-find proceed exactly as in the CIR case.
\end{remark}

\begin{remark}
\label{rmk:jacobi-note}
Another analytically tractable model can be obtained by setting the functions $(r,a,b,c)$ equal to
\begin{align}
r(y)
	&=	\frac{\eta y}{1-y} , &
b(y)
	&=	\kappa - \theta y  , &
a(y)
	&=	\del \sqrt{y(1-y)} , &
c(y)
	&=	\gam \sqrt{\frac{y}{1-y}} .
\end{align}
In this case, however, the volatility of $X$ is proportional to the square root of the interest rate: $c(Y) = \gam \sqrt{R/\eta}$.
\end{remark}

%
%

\section{Conclusion}
\label{sec:conclusion}
This paper develops a flexible class of short‑rate–dependent volatility models in which the instantaneous volatility of a risky asset is an explicit function of the driver $Y$ of the short rate. Within this framework, European option prices are expressed via a Fourier representation that reduces the pricing problem to the computation of a characteristic function $G$ associated with the time-integrals of the short rate $r(Y)$ and stochastic variance $c^2(Y)$. The derivation is carried out entirely at the level of partial differential equations: the option value solves a backward pricing equation, a Fourier transform in the log-price variable removes that variable and leaves a one-dimensional parabolic equation for $G$ with a complex first-order coefficient, and inverting the transform yields the price.  The complex coefficient enters as an ordinary (if complex) PDE coefficient and is interpreted, in the examples, as the analytic continuation in the model parameters of a real-parameter transform; no change of probability measure is required.
\\[0.5em]
Two analytically tractable specifications illustrate the approach: a CIR driver $Y$ with a short rate equal to $Y$ and a volatility proportional to $1/\sqrt{Y}$, and a Jacobi driver $Y$ with a short rate proportional to $Y/(1-Y)$ and a volatility proportional to $\sqrt{(1-Y)/Y}$.  In both cases, explicit formulas for the relevant Laplace transforms, and hence for the characteristic function $G$, are available through results of \cite{hurd2008explicit}, which in turn yield closed-form integral expressions for European call prices from which the associated Black-Scholes implied volatilities can be computed numerically. Numerical illustrations highlight the ability of short-rate-dependent volatility to generate asymmetric implied volatility smiles, even in parameter regimes where the covariance between the log-price and the short rate vanishes under the original pricing measure.
\\[0.5em]
The framework can be extended in several directions, including multi-factor specifications for the short rate, alternative functional forms for the volatility-rate linkage, and the valuation of path‑dependent or exotic claims under the same Fourier transform methodology. Such extensions would allow for a more detailed study of how monetary policy and interest rate uncertainty are transmitted into the equity volatility surface. Therefore, they may help bridge the gap between reduced-form stochastic volatility models and term-structure models used in fixed-income markets.

%
%

\appendix

\section{Dynamics of $F^T$ under the $T$-forward measure in the CIR setting}
\label{sec:T-forward}
In this Appendix, we derive the dynamics of $(Z^T = \log F^T,Y)$ under the $T$-forward probability measure for the model presented in Section \ref{sec:cir}.
To begin, we note that the dynamics of the price of the risky asset $S$ and the price of a zero-coupon bond $B^T$ can be written as
\begin{align}
\dd S_t
	&=	r(Y_t) S_t \dd t + c(Y_t) S_t ( \rho \dd W_t + \rhob \dd B_t ) , &
B_t^T
	&=	\exp \Big( - P(t;T) - Q(t;T) Y_t \Big ) , 
\end{align}
where the functions $P$ and $Q$ are given in \eqref{eq:PQ}.
From this, we compute the dynamics of $S/M$ and $B^T/M$, both of which are martingales under $\Pb$.  We have
\begin{align}
\dd \Big( \frac{S_t}{M_t} \Big)
	&= c(Y_t) \Big( \frac{S_t}{M_t} \Big) ( \rho \dd W_t + \rhob \dd B_t ) , &
\dd \Big( \frac{B_t^T}{M_t} \Big)
	&=	- a(Y_t) Q(t;T) \Big( \frac{B_t^T}{M_t} \Big) \dd W_t .
\end{align}
Next, we derive the dynamics of $F^T = S/B^T$, which is given by
\begin{align}
\dd F_t^T
	&=	\dd \Big( \frac{S_t}{B_t^T} \Big)
	=		\dd \Big( \frac{S_t/M_t}{B_t^T/M_t} \Big) \\
	&=	\Big( a^2(Y_t) Q^2(t;T) + \rho a(Y_t) Q(t;T) c(Y_t) \Big) F_t^T \dd t \\ &\quad
			+ \Big( a(Y_t) Q(t;T) + \rho c(Y_t) \Big) F_t^T \dd W_t + \rhob c(Y_t) F_t^T \dd B_t . 
\end{align}
Now, let us define the \textit{$T$-forward probability measure} $\Pbh$, whose relation to $\Pb$ is given by the following change of measure
\begin{align}
\frac{\dd \Pbh}{\dd \Pb}
	&:=	\frac{B_T^T M_0}{B_0^T M_T}
	=		\exp \Big( - \frac{1}{2} \int_0^T (a(Y_t) Q(t;T) )^2 \dd t - \int_0^T \big( a(Y_t) Q(t;T) \big) \dd W_t \Big) .
\end{align}
Under $\Pbh$, the processes $\Wh$ and $\Bh$, defined by
\begin{align}
\dd \Wh_t
	&=	\dd W_t + a(Y_t) Q(t;T) \dd t , &
\Wh_0
	&=	0 , \\
\dd \Bh_t
	&=	\dd B_t , &
\Bh_0
	&=	0 , &
\end{align}
are independent Brownian motion.  As such, the dynamics of $F^T$ under $\Pbh$ are
\begin{align}
\dd F_t^T
	&=	\Big( a^2(Y_t) Q^2(t;T) + \rho a(Y_t) Q(t;T) c(Y_t) \Big) F_t^T \dd t \\ &\quad
			+ \Big( a(Y_t) Q(t;T) + \rho c(Y_t) \Big) F_t^T ( \dd \Wh_t - a(Y_t) Q(t;T) \dd t ) + \rhob c(Y_t) F_t^T \dd \Bh_t , \\
	&=	\Big( a(Y_t) Q(t;T) + \rho c(Y_t) \Big) F_t^T \dd \Wh_t  + \rhob c(Y_t) F_t^T \dd \Bh_t .
\end{align}
Observe that $F^T$ is a martingale under the $T$-forward measure $\Pbh$, as it should be.
Finally, the dynamics of $Y$ and $Z^T:= \log F^T$ under $\Pbh$ are
\begin{align}
\dd Y_t
	&=	b(Y_t) \dd t + a(Y_t) ( \dd \Wh_t - a(Y_t) Q(t;T) \dd t  )  \\
	&=	\Big( b(Y_t) - a^2(Y_t) Q(t;T) \Big) \dd t + a(Y_t) \dd \Wh_t , \label{eq:dY-Tforward} \\
\dd Z_t^T
	&=	- \frac{1}{2} \Big( a^2(Y_t) Q^2(t;T) + c^2(Y_t) + 2 \rho a(Y_t) Q(t;T) c(Y_t) \Big) \dd t \\ &\quad
			+ \Big( a(Y_t) Q(t;T) + \rho c(Y_t) \Big) \dd \Wh_t  + \rhob c(Y_t) \dd \Bh_t . \label{eq:dZ-Tforward}
\end{align}

%
%

\section{\textit{Wolfram Mathematica} \textcopyright{} code}
\label{sec:mathematica}
In this Appendix, we provide the \textit{Wolfram Mathematica} \textcopyright{} code used to produce the implied volatility plots.  Appendix~\ref{sec:mathematica-cir} contains the code for the CIR-driven model of Section~\ref{sec:cir} (Figure~\ref{fig:cir-iv}), and Appendix~\ref{sec:mathematica-jacobi} contains the code for the Jacobi-driven model of Section~\ref{sec:jacobi} (Figure~\ref{fig:jacobi-iv}).  In each case, evaluating the single code cell defines all functions and produces the two implied-volatility plots.

\subsection{CIR-driven short rate}
\label{sec:mathematica-cir}
\begin{lstlisting}
ClearAll["Global`*"]

(* ============================================================ *)
(* Implied volatility for the CIR-driven short-rate model.       *)
(*   r(y)=y,  b(y)=kappa(theta-y),                               *)
(*   a(y)=delta Sqrt[y],  c(y)=gamma/Sqrt[y].                    *)
(* G is the Hurd-Kuznetsov CIR transform (Theorem 3.1),          *)
(* evaluated at the complex mean-reversion level thetaT          *)
(* generated by the Fourier variable.                           *)
(* ============================================================ *)

(* G^CIR : Hurd-Kuznetsov CIR transform (their Theorem 3.1).     *)
(* Arguments: maturity T, state x(=y), transform weights d1,d2,  *)
(* offsets w1,w2, and CIR parameters a=kappa thetaT, b=kappa,    *)
(* c=delta.                                                      *)
GCIR[T_, x_, d1_, d2_, w1_, w2_, a_, b_, c_] := Module[
   {al, be, v1, v2, gT, zArg, exponentTerm, preFactor, hyperGeomTerm},
   al = 2 a/c^2 - 1;
   be = 2 b/c^2;
   v1 = (1/2) (-be + Sqrt[be^2 + 8 d1/c^2]);
   v2 = (1/2) (-al + Sqrt[al^2 + 8 d2/c^2]);
   gT = (be + 2 v1)/(1 - Exp[-(be/2 + v1) c^2 T]);
   zArg = -((gT^2 x Exp[-(be/2 + v1) c^2 T])/(gT + w1 - v1));
   preFactor = Exp[-(a v1 + b v2 + c^2 v1 v2) T] x^v2*
     (gT + w1 - v1)^(-al - v2 - w2 - 1) gT^(al + 2 v2 + 1);
   exponentTerm = Exp[-x (v1 + (gT (w1 - v1))/(gT + w1 - v1)*
        Exp[-(be/2 + v1) c^2 T])];
   hyperGeomTerm = (Gamma[al + v2 + w2 + 1]/Gamma[al + 2 v2 + 1])*
     Hypergeometric1F1[v2 - w2, al + 2 v2 + 1, zArg];
   preFactor exponentTerm hyperGeomTerm
   ];

(* Map the CIR model onto GCIR.  The Fourier transform makes the *)
(* mean-reversion level complex: thetaT = theta + I w rho delta  *)
(* gamma/kappa.  Here ww,zz are the transform weights (w,z).      *)
GJac[w_, T_, y_, ww_, zz_, kappa_, theta_, delta_, gamma_, rho_] :=
  GCIR[T, y, ww, gamma^2 zz, 0, 0,
   kappa (theta + I w rho delta gamma/kappa), kappa, delta];

(* Zero-coupon bond B(t,y;T) = G at omega=0, ww=1, zz=0.          *)
bondCIR[T_, y_, kappa_, theta_, delta_, gamma_] :=
  Re[GJac[0, T, y, 1, 0, kappa, theta, delta, gamma, 0]];

(* Generalized Fourier transform of the call payoff.             *)
PhiHat[w_, K_] := -K^(1 - I w)/(w^2 + I w);

(* Spot call price via Fourier inversion (finite range, machine  *)
(* precision: the integrand decays quickly in wr).               *)
callPrice[T_, x_, y_, kappa_, theta_, delta_, gamma_, rho_, K_] :=
  Module[{wi, w, ww, zz},
   wi = -21/20;                      (* contour Im(omega) in (-1.15,-1) *)
   w = wr + I wi;
   ww = 1 - I w;
   zz = w^2/2 + I w/2;
   (1/Pi) NIntegrate[
     Re[PhiHat[w, K] Exp[I w x]*
        GJac[w, T, y, ww, zz, kappa, theta, delta, gamma, rho]],
     {wr, 0, 50}, WorkingPrecision -> MachinePrecision, MaxRecursion -> 12]
   ];

(* Black-Scholes call (on the forward).                          *)
blackScholesCall[F_, K_, tau_, sigma_] := Module[{d1, d2},
   d1 = (Log[F/K] + (sigma^2/2) tau)/(sigma Sqrt[tau]);
   d2 = d1 - sigma Sqrt[tau];
   F CDF[NormalDistribution[], d1] - K CDF[NormalDistribution[], d2]];

(* Implied volatility for given log-moneyness L and correlation. *)
impliedVol[T_, x_, y_, kappa_, theta_, delta_, gamma_, rho_, L_] :=
  Module[{B, K, cModel, cForward, F},
   B = bondCIR[T, y, kappa, theta, delta, gamma];
   K = Exp[x + L]/B;
   F = Exp[x]/B;
   cModel = callPrice[T, x, y, kappa, theta, delta, gamma, rho, K];
   cForward = cModel/B;
   If[cForward <= 0 || cForward >= F, Return[0]];
   sigma /. Quiet@FindRoot[blackScholesCall[F, K, T, sigma] == cForward,
      {sigma, 0.25}, MaxIterations -> 100,
      WorkingPrecision -> MachinePrecision]];

(* ============================ Parameters ============================ *)
kappa = 1/2;      (* CIR mean-reversion speed                            *)
theta = 1/25;     (* long-run short rate = 0.04                          *)
delta = 9/50;     (* vol-of-rate = 0.18; Feller 2 kappa theta > delta^2  *)
gamma = 1/20;     (* vol scale: c(y0)=gamma/Sqrt[y0]=0.25 at y0=0.04     *)
t = 0;
x = Log[100];     (* current log-price, S0 = 100                         *)
y = 1/25;         (* initial short rate Y0 = R0 = 0.04, vol0 = 0.25      *)

(* ============================ Plots ============================ *)
Print["Implied vol vs log-moneyness, T = 0.25"]
Lmin = -0.12; Lmax = 0.12; T = 0.25;
Quiet[Plot[{
   impliedVol[T, x, y, kappa, theta, delta, gamma, 1/4, L],
   impliedVol[T, x, y, kappa, theta, delta, gamma, 0, L],
   impliedVol[T, x, y, kappa, theta, delta, gamma, -1/4, L]
   }, {L, Lmin, Lmax}, PlotPoints -> 25, MaxRecursion -> 0,
  PlotStyle -> {{Black, Dotted}, {Black, Dashed}, Black},
  AxesLabel -> {"L", "\[Sigma]"}, PlotRange -> All]]

Print["Implied vol vs log-moneyness, T = 0.50"]
Lmin = -0.16; Lmax = 0.16; T = 0.5;
Quiet[Plot[{
   impliedVol[T, x, y, kappa, theta, delta, gamma, 1/4, L],
   impliedVol[T, x, y, kappa, theta, delta, gamma, 0, L],
   impliedVol[T, x, y, kappa, theta, delta, gamma, -1/4, L]
   }, {L, Lmin, Lmax}, PlotPoints -> 25, MaxRecursion -> 0,
  PlotStyle -> {{Black, Dotted}, {Black, Dashed}, Black},
  AxesLabel -> {"L", "\[Sigma]"}, PlotRange -> All]]
\end{lstlisting}

\subsection{Jacobi-driven short rate}
\label{sec:mathematica-jacobi}
\begin{lstlisting}
ClearAll["Global`*"]

(* ============================================================ *)
(* Implied volatility for the JACOBI-driven short-rate model.    *)
(*   r(y)=eta y/(1-y),  b(y)=kappa-theta y,                       *)
(*   a(y)=delta Sqrt[y(1-y)],  c(y)=gamma Sqrt[(1-y)/y].          *)
(* G is the Hurd-Kuznetsov Jacobi transform (Theorem 4.1),       *)
(* evaluated at the complex parameters (kappaT,thetaT) generated  *)
(* by the Fourier variable.                                      *)
(* ============================================================ *)

(* Jacobi polynomial via its 2F1 representation, valid for       *)
(* COMPLEX upper parameters a,b (Mathematica's built-in JacobiP  *)
(* is unreliable for complex indices).                           *)
JacobiPc[n_, a_, b_, x_] := Pochhammer[a + 1, n]/n! *
   Hypergeometric2F1[-n, n + a + b + 1, a + 1, (1 - x)/2];

(* G^Jacobi : solution of the G-PDE for the Jacobi model.        *)
(* Arguments: Fourier var w (=omega), maturity T, state y,        *)
(* transform args ww,zz, and model params.                       *)
GJacobi[w_, T_, y_, ww_, zz_, kappa_, theta_, delta_, eta_, gamma_, rho_,
   nMax_ : 50] := Module[
  {kappaT, thetaT, al, be, v1, v2, A, pre, series},
  kappaT = kappa + I w rho delta gamma;
  thetaT = theta + I w rho delta gamma;
  al = 2 kappaT/delta^2 - 1;
  v1 = (-al + Sqrt[al^2 + 8 zz gamma^2/delta^2])/2;
  be = 2 (thetaT - kappaT)/delta^2 - 1;
  v2 = (-be + Sqrt[be^2 + 8 ww eta/delta^2])/2;
  A = al + be + 2 v1 + 2 v2 + 1;
  pre = Exp[-((thetaT - kappaT) v1 + kappaT v2 + delta^2 v1 v2) T]*
    y^v1 (1 - y)^v2*
    (Gamma[al + v1 + 1] Gamma[be + v2 + 1] Gamma[al + be + 2 v1 + 2 v2 + 1])/
    (Gamma[al + v1 + be + v2 + 2] Gamma[al + 2 v1 + 1] Gamma[be + 2 v2 + 1]);
  series = Sum[
    Exp[-nn (nn + A) delta^2 T/2]*
      Pochhammer[A, nn]/Pochhammer[al + 2 v1 + 1, nn]*(2 nn + A)*
      HypergeometricPFQ[{-nn, A + nn, be + v2 + 1},
        {al + be + v1 + v2 + 2, be + 2 v2 + 1}, 1]*
      JacobiPc[nn, be + 2 v2, al + 2 v1, 2 y - 1],
    {nn, 0, nMax}];
  pre series
];

(* Zero-coupon bond B(t,y;T) = G at omega=0, ww=1, zz=0.          *)
bondJacobi[T_, y_, kappa_, theta_, delta_, eta_, gamma_] :=
  Re[GJacobi[0, T, y, 1, 0, kappa, theta, delta, eta, gamma, 0]];

(* Generalized Fourier transform of the call payoff.             *)
PhiHat[w_, K_] := -K^(1 - I w)/(w^2 + I w);

(* Spot call price via Fourier inversion.                        *)
callPrice[T_, x_, y_, kappa_, theta_, delta_, eta_, gamma_, rho_, K_] :=
  Module[{wi, w, ww, zz},
   wi = -3/2;                       (* contour Im(omega) < -1     *)
   w = wr + I wi;
   ww = 1 - I w;
   zz = w^2/2 + I w/2;
   (1/Pi) NIntegrate[
     Re[PhiHat[w, K] Exp[I w x]*
        GJacobi[w, T, y, ww, zz, kappa, theta, delta, eta, gamma, rho]],
     {wr, 0, 50}, WorkingPrecision -> MachinePrecision, MaxRecursion -> 12]
   ];

(* Black-Scholes call (on the forward).                          *)
blackScholesCall[F_, K_, tau_, sigma_] := Module[{d1, d2},
   d1 = (Log[F/K] + (sigma^2/2) tau)/(sigma Sqrt[tau]);
   d2 = d1 - sigma Sqrt[tau];
   F CDF[NormalDistribution[], d1] - K CDF[NormalDistribution[], d2]];

(* Implied volatility for given log-moneyness L and correlation. *)
impliedVol[T_, x_, y_, kappa_, theta_, delta_, eta_, gamma_, rho_, L_] :=
  Module[{B, K, cModel, cForward, F},
   B = bondJacobi[T, y, kappa, theta, delta, eta, gamma];
   K = Exp[x + L]/B;
   F = Exp[x]/B;
   cModel = callPrice[T, x, y, kappa, theta, delta, eta, gamma, rho, K];
   cForward = cModel/B;
   If[cForward <= 0 || cForward >= F, Return[0]];
   sigma /. Quiet@FindRoot[blackScholesCall[F, K, T, sigma] == cForward,
      {sigma, 0.25}, MaxIterations -> 100, WorkingPrecision -> MachinePrecision]];

(* ============================ Parameters ============================ *)
kappa = 1/2;     (* Jacobi mean-reversion speed                         *)
theta = 1;     (* so stationary mean kappa/theta = 0.5                *)
delta = 7/10;     (* vol-of-Y; keeps Y in (0,1): kappa,theta-kappa>delta^2/2 *)
eta = 1/25;    (* rate scale: r(y0)=eta y0/(1-y0)=0.04 at y0=0.5       *)
gamma = 1/4;    (* vol scale: c(y0)=gamma Sqrt[(1-y0)/y0]=0.25 at y0=0.5 *)
t = 0;
x = Log[100];    (* current log-price, S0 = 100                         *)
y = 1/2;         (* initial driver Y0 = 0.5  =>  R0 = 0.04, vol0 = 0.25  *)

(* ============================ Plots ============================ *)
Print["Implied vol vs log-moneyness, T = 0.25"]
Lmin = -0.12; Lmax = 0.12; T = 0.25;
Quiet[Plot[{
   impliedVol[T, x, y, kappa, theta, delta, eta, gamma, 1/4, L],
   impliedVol[T, x, y, kappa, theta, delta, eta, gamma, 0, L],
   impliedVol[T, x, y, kappa, theta, delta, eta, gamma, -1/4, L]
   }, {L, Lmin, Lmax}, PlotPoints -> 25, MaxRecursion -> 0,
  PlotStyle -> {{Black, Dotted}, {Black, Dashed}, Black},
  AxesLabel -> {"L", "\[Sigma]"}, PlotRange -> All]]

Print["Implied vol vs log-moneyness, T = 0.50"]
Lmin = -0.16; Lmax = 0.16; T = 0.5;
Quiet[Plot[{
   impliedVol[T, x, y, kappa, theta, delta, eta, gamma, 1/4, L],
   impliedVol[T, x, y, kappa, theta, delta, eta, gamma, 0, L],
   impliedVol[T, x, y, kappa, theta, delta, eta, gamma, -1/4, L]
   }, {L, Lmin, Lmax}, PlotPoints -> 25, MaxRecursion -> 0,
  PlotStyle -> {{Black, Dotted}, {Black, Dashed}, Black},
  AxesLabel -> {"L", "\[Sigma]"}, PlotRange -> All]]
\end{lstlisting}

%
%

\clearpage
\bibliography{references-2}

%
%

\clearpage

\begin{figure}
\begin{tabular}{cc}
\includegraphics[width=0.475\textwidth]{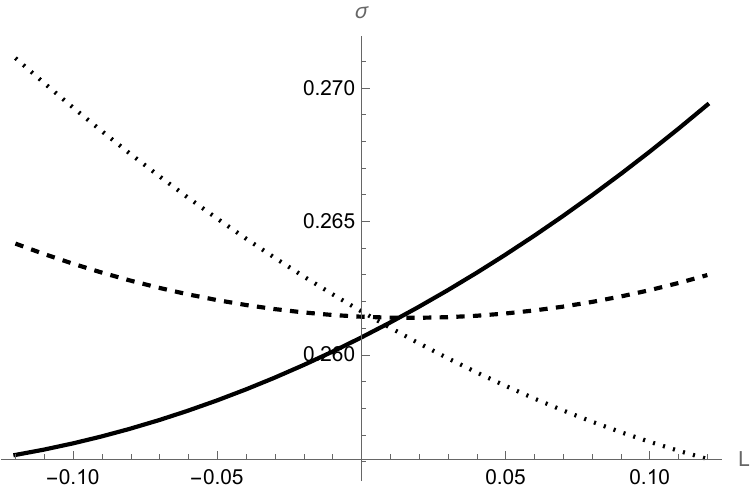}&
\includegraphics[width=0.475\textwidth]{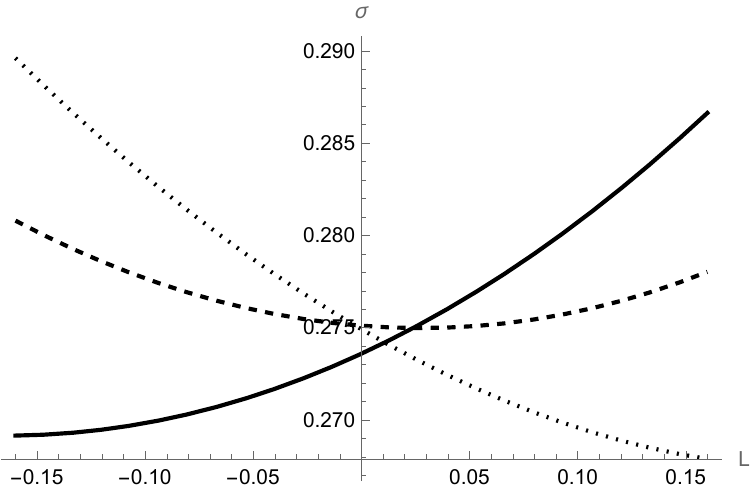}\\
$T=0.25$ & $T=0.50$
\end{tabular}
\caption{{For the model described in Section \ref{sec:cir}, implied volatility $\sig_0(T,L)$ is plotted as a function of log-moneyness $L$ for two different maturities, $T=0.25$ and $T=0.50$, and three correlation coefficients, $\rho = 0.25$ (dotted), $\rho = 0.0$ (dashed) and $\rho = -0.25$ (solid). Other parameters are fixed at the following values: $\kappa = 0.5$, $\theta = 0.04$, $\del = 0.18$, $\gamma = 0.05$, $t=0$, $X_0 = \log 100$, and $Y_0 = \theta = 0.04$. These values give an initial short rate $R_0 = Y_0 = 0.04$ and an initial volatility $c(Y_0) = \gamma/\sqrt{Y_0} = 0.25$, and satisfy the Feller condition $2\kappa\theta > \del^2$.}}
\label{fig:cir-iv}
\end{figure}

\begin{figure}
\begin{tabular}{cc}
\includegraphics[width=0.475\textwidth]{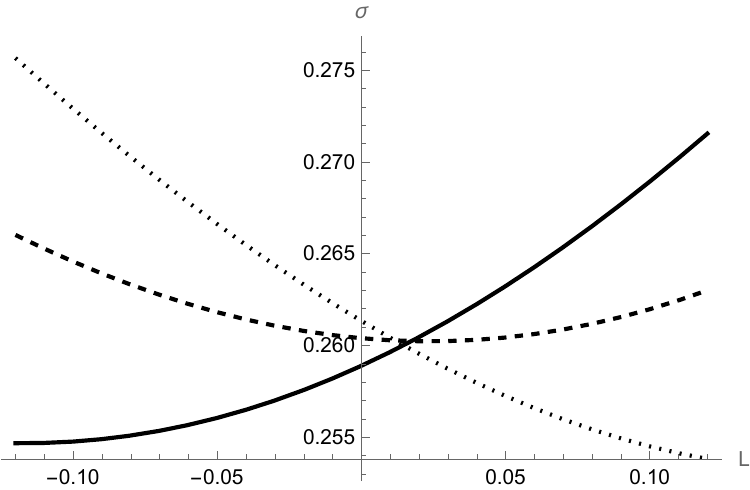}&
\includegraphics[width=0.475\textwidth]{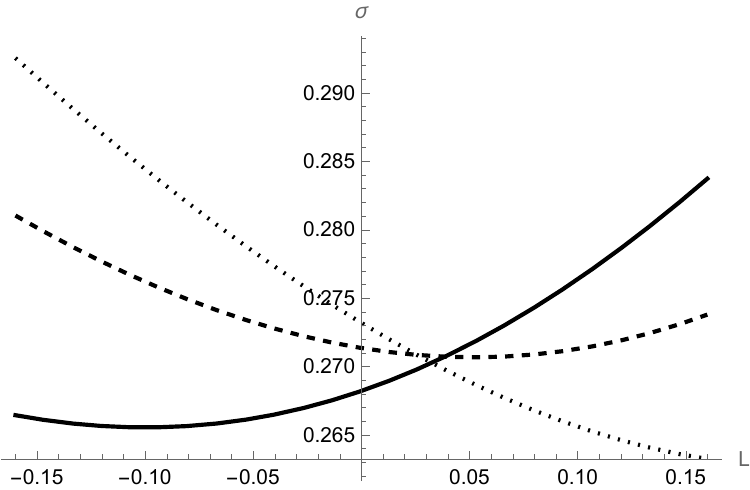}\\
$T=0.25$ & $T=0.50$
\end{tabular}
\caption{{For the model described in Section \ref{sec:jacobi}, implied volatility $\sig_0(T,L)$ is plotted as a function of log-moneyness $L$ for two different maturities, $T=0.25$ and $T=0.50$, and three correlation coefficients, $\rho = 0.25$ (dotted), $\rho = 0.0$ (dashed) and $\rho = -0.25$ (solid). Other parameters are fixed at the following values: $\kappa = 0.5$, $\theta = 1.0$, $\del = 0.7$, $\eta = 0.04$, $\gamma = 0.25$, $t=0$, $X_0 = \log 100$, and $Y_0 = 0.5$. These values give an initial short rate $R_0 = \eta Y_0/(1-Y_0) = 0.04$ and an initial volatility $c(Y_0) = \gamma\sqrt{(1-Y_0)/Y_0} = 0.25$, and satisfy the conditions $\kappa > \del^2/2$ and $\theta - \kappa > \del^2/2$ that keep $Y$ in $(0,1)$.}}
\label{fig:jacobi-iv}
\end{figure}

\end{document}